\documentclass[aps,prb,twocolumn,showpacs,superscriptaddress]{revtex4}

\usepackage{graphicx}
\usepackage{dcolumn}
\usepackage{bm}
\usepackage{stmaryrd}
\usepackage{latexsym}
\usepackage{amssymb}
\usepackage{amsfonts}
\usepackage{amsmath}
\usepackage{fancybox}
\usepackage{verbatim}
\usepackage{ulem}

\begin{document}

\title{Electron-electron interactions in decoupled graphene layers}

\author{Rosario E.V. Profumo}
\affiliation{NEST, Istituto Nanoscienze-CNR and Dipartimento di Fisica dell'Universit\`a di Pisa, I-56127 Pisa, Italy}
\author{Marco Polini}
\email{m.polini@sns.it}
\affiliation{NEST, Istituto Nanoscienze-CNR and Scuola Normale Superiore, I-56126 Pisa, Italy}
\author{Reza Asgari}
\affiliation{School of Physics, Institute for Research in Fundamental Sciences (IPM), Tehran 19395-5531, Iran}
\author{Rosario Fazio}
\affiliation{NEST, Scuola Normale Superiore and Istituto Nanoscienze-CNR, I-56126 Pisa, Italy}
\author{A.H. MacDonald}
\affiliation{Department of Physics, The University of Texas at Austin, Austin, Texas 78712, USA}

\begin{abstract}
Multi-layer graphene on the carbon face of silicon carbide is an intriguing electronic system which 
typically consists of a stack of ten or more layers. Rotational stacking faults in this system 
dramatically reduce inter-layer coherence.  In this article we report on the influence of 
inter-layer interactions, which remain strong even when coherence is negligible, on the Fermi liquid 
properties of charged graphene layers.  We find that inter-layer interactions increase the magnitudes of correlation energies and 
decrease quasiparticle velocities, even when remote-layer carrier densities are small,
and that they lessen the influence of exchange and correlation on the distribution 
of carriers across layers.  
\end{abstract}

\pacs{73.21.Ac,81.05.ue,73.22.Pr,71.10.-w}

\maketitle

\section{Introduction}
\label{sect:intro}

Graphene layers prepared by the thermal decomposition of silicon carbide (SiC)~\cite{berger_science_2006,rollings_jpcs_2006,deheer_ssc_2007,bostwick_natphys_2007,zhou_natmater_2007,emtsev_prb_2008,emtsev_natmater_2009,riedl_prl_2009,first_arxiv_2010} might play
a pivotal role in carbon-based analog electronics because of their high carrier mobilities, and because of the possibility of 
direct growth on a well understood single-crystal semiconductor substrate.   
This form of graphene, usually termed {\it epitaxial graphene}, can be grown directly on either 
silicon or carbon terminated faces of SiC and is scalable to large area circuits. 

Epitaxial graphene on the carbon face typically consists of tens of graphene layers (see however Ref.~\onlinecite{wu_apl_2009}), 
with rotational stacking faults that lead to an electronic structure which is practically 
indistinguishable~\cite{sadowski_prl_2006,wu_prl_2007,santos_prl_2007,varchon_prl_2007,hass_prl_2008,orlita_prl_2008,
varchon_prb_2008,sprinkle_prl_2009} from that of a sample with many electrically isolated single-layer graphene (SLG)~\cite{castroneto_rmp_2009} layers. It is commonly believed~\cite{berger_science_2006,rollings_jpcs_2006,deheer_ssc_2007,bostwick_natphys_2007,zhou_natmater_2007,emtsev_prb_2008,emtsev_natmater_2009,riedl_prl_2009,first_arxiv_2010,sadowski_prl_2006,varchon_prl_2007,orlita_prl_2008} that the layer closest to the SiC 
typically has a rather high carrier density $\approx 10^{12}~{\rm cm}^{-2}$, due to charge transfer from the substrate, and that 
the overlayers are nearly neutral and unimportant for most observables.    
Even though multi-layer graphene (MLG) on the C-face of SiC has been shown to possess many of the distinctive features of SLG, 
including the half-quantized quantum Hall effect~\cite{tzalenchuk_natnano_2010,wu_apl_2009}, two key questions arise: 
i) to what extent do Coulomb interactions between electrons in nearby layers distinguish the electronic properties of MLG 
layers from those of SLG? ii) what is the role of inter-layer Coulomb interactions in determining the electron-density distribution
which achieves equilibrium between layers?  These are the main issues we address in this article. 

%%%%%%%%%%%%%%
\begin{figure}[t]
\centering
\includegraphics[width=0.60\linewidth]{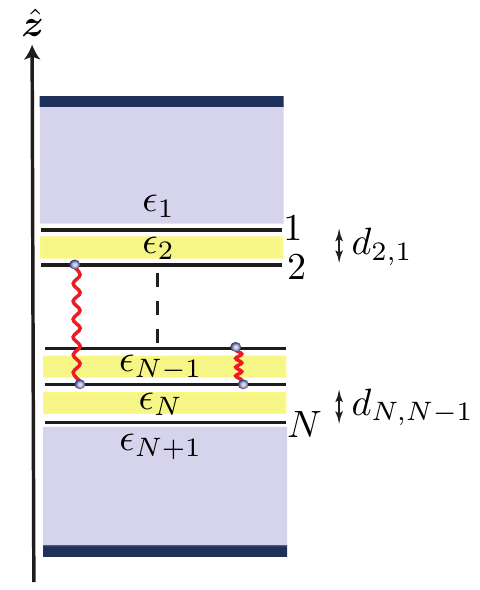}
\caption{(Color online) A stack of $\ell = 1 \dots N$ graphene layers is placed between top and bottom dielectrics. 
The layers are coupled solely by inter-layer Coulomb interactions (red wiggly lines). The dielectric constant in the region above the $\ell$-th layer is labeled by $\epsilon_{\ell}$, while the one of the substrate is labeled by $\epsilon_{N+1}$. The separation between the $\ell+1$-th and the $\ell$-th layer is labeled by $d_{\ell+1,\ell}$.\label{fig:one}}
\end{figure}
%%%%%%%%%%%%%%

The system we study is sketched in Fig.~\ref{fig:one}. We assume that the graphene layers are perfectly 
decoupled from the point of view of single-particle tunneling, but take Coulomb interactions between electrons in different layers, which provide a source of (two-particle) coupling,
fully into account.  We study, (i) the impact of intra- and inter-layer Coulomb interactions on the Fermi-liquid properties of MLG two-dimensional 
electron systems, and (ii) the combined role of Hartree, exchange, and correlation potentials in determining the equilibrium distribution of carrier densities across layers when charge is transferred to the many-layer system from the SiC substrate. The theory sketched below applies to MLG on the C-face of SiC with an arbitrary number $N$ of layers, to the decoupled layers sometimes found on the surface of bulk graphite~\cite{grapheneongraphite}, and to the experiments by Schmidt {\it et al.}~\cite{schmidt_prb_2010} on micromechanically exfoliated decoupled layers.  For the sake of definiteness and simplicity, we present explicit numerical results only for $n$-type carriers and for the
$N=2$ (double-layer) case.

This manuscript is organized as follows. In Sect.~\ref{sect:model} we present the model we have used to describe tunnel-decoupled 
graphene layers in the presence of intra- and inter-layer electron-electron interactions, and discuss the linear-response functions which control the physical properties in which we are interested. In Sect.~\ref{sect:mainresults} we describe the minimal microscopic theory that allows us to calculate quasiparticle velocities, ground-state energies, and chemical potentials. In Sect.~\ref{sect:electrochemicalequilibrium} we present a general scheme to tackle the electro-chemical equilibrium problem in MLG. Finally, in Sect.~\ref{sect:numericalresults} we present our main numerical results for double-layer graphene. Sect.~\ref{sect:summary} contains a summary of our main conclusions. Appendix~\ref{appendix} collects some lengthy expressions for the exchange and correlation energies of double-layer graphene that are important for the technical details of our calculations.

\section{Model Hamiltonian and linear-response theory}
\label{sect:model}

Our model Hamiltonian contains massless-Dirac-fermion kinetic-energy terms and intra- and inter-layer Coulomb interactions: 
\begin{eqnarray}\label{eq:Hamiltonian}
{\hat {\cal H}} &=&  \hbar v \sum_{{\bm k}, \ell, \alpha, \beta} {\hat \psi}^\dagger_{{\bm k}, \ell, \alpha} 
( {\bm \sigma}_{\alpha\beta} \cdot {\bm k} ) {\hat \psi}_{{\bm k}, \ell, \beta} \nonumber\\
&+& \frac{1}{2 S}\sum_{{\bm q} \neq 0, \ell, \ell'} V_{\ell \ell'}(q){\hat \rho}_{{\bm q}, \ell} {\hat \rho}_{-{\bm q}, \ell'}~.
\end{eqnarray}
Here $v$ is the bare Fermi velocity, taken to be the same in all the $\ell = 1 \dots N$ tunnel-decoupled layers, $S$ is the area of each layer, $V_{\ell \ell'}(q)$ is the matrix of bare Coulomb potentials, and 
\begin{equation}\label{eq:densityoperator}
{\hat \rho}_{{\bm q}, \ell} = \sum_{{\bm k}, \alpha} {\hat \psi}^\dagger_{{\bm k} - {\bm q}, \ell, \alpha}{\hat \psi}_{{\bm k}, \ell, \alpha}
\end{equation}
is the density-operator for the $\ell$-th layer. 
Greek letters are honeycomb-sublattice-pseudospin labels and ${\bm \sigma} = (\sigma^x,\sigma^y)$ is a vector of Pauli matrices. 
We also introduce the coupling constant~\cite{castroneto_rmp_2009} $\alpha_{\rm ee} = e^2/(\hbar v)$, whose value is $\approx 2.2$ if for $v$ we use the SLG Fermi velocity $v_{\rm F} \approx 10^{6}~{\rm m}/{\rm s}$.

Several important many-body properties of the Hamiltonian ${\hat {\cal H}}$ are completely determined by the $N \times N$ 
symmetric matrix ${\bm \chi}(q,\omega)$ whose elements are the density-density linear-response functions
\begin{equation}\label{eq:LRT}
\chi_{\ell\ell'}(q,\omega) = \frac{1}{S} \langle \langle {\hat \rho}_{{\bm q}, \ell}; {\hat \rho}_{-{\bm q}, \ell}\rangle\rangle_\omega~,
\end{equation}
with $\langle\langle {\hat A},{\hat B}\rangle\rangle_\omega$ the usual Kubo product~\cite{Giuliani_and_Vignale}.  Within the random phase approximation (RPA) these functions satisfy the following matrix equation,
\begin{equation}\label{eq:matrix-form}
{\bm \chi}^{-1}(q,\omega) = {\bm \chi}^{-1}_0(q,\omega) - {\bm V}(q)~,
\end{equation}
where ${\bm \chi}_0(q,\omega)$ is a $N \times N$ diagonal matrix whose elements $\chi^{(0)}_\ell(q,\omega)$ are the well-known~\cite{barlas_prl_2007,wunsch_njp_2006,hwang_prb_2007} noninteracting (Lindhard) response functions of each graphene layer at arbitrary doping $n_\ell = N_\ell/S$.  The off-diagonal (diagonal) elements of the matrix ${\bm V} =\{V_{\ell \ell'}\}_{\ell,\ell' = 1 \dots N}$ represent inter-layer (intra-layer) Coulomb interactions. 

\section{Quasiparticle velocities, ground-state energy, and chemical potentials}
\label{sect:mainresults}

The Fermi-liquid parameters of MLG can be calculated from the knowledge of the retarded quasiparticle 
self-energy $\Sigma_\ell$. Our results are based on the so-called ``${\rm G}_0{\rm W}$" approximation~\cite{Giuliani_and_Vignale,hedin_pr_1965,asgari_prb_2005,polini_ssc_2007,hwang_arpes_2008} in which the self-energy $\Sigma_\ell$ is expanded to first order in the dynamically screened effective interaction ${\bm W}$~\cite{Giuliani_and_Vignale},
\begin{eqnarray}\label{eq:e-e}
{\bm W}(q,\omega) &= & {\bm V}(q)+{\bm V}(q){\bm \chi}(q,\omega){\bm  V}(q) \nonumber \\
&=&  [{\bm V}^{-1}(q) -{\bm \chi}_0(q,\omega)]^{-1}~.
\end{eqnarray}
In this paper we limit our attention to many-body quantities that can be expressed solely in terms of integrals along the imaginary frequency axis where the Lindhard function has a smooth dependence on its parameters. 

The microscopic expression for $\Sigma_\ell$ in terms of ${\bm W}$ can be obtained by a straightforward 
generalization of the theory of Ref.~\onlinecite{polini_ssc_2007} to a multicomponent system ($\hbar=1$):
\begin{eqnarray}\label{eq:sigmaell}
\Sigma_{\ell}({\bm k},i\omega_n) &=& - \frac{1}{\beta}\sum_{s}
\int \frac{d^2{\bm q}}{(2\pi)^2} \sum_{m=-\infty}^{+\infty}
W_{\ell\ell}(q,i\Omega_m) \nonumber\\ 
&\times &\left[\frac{1+s\cos{(\theta_{{\bm k},{\bm k}'})}}{2}\right] G^{(0)}_{\ell, s}({\bm k}',i\omega'_{n, m})~,
\end{eqnarray}
where $\beta = (k_{\rm B} T)^{-1}$, ${\bm k}' = {\bm k}+{\bm q}$ and $\omega'_{n, m} =\omega_n + \Omega_m$. In Eq.~(\ref{eq:sigmaell}) $\omega_n=(2n+1)\pi/\beta$ is a fermionic Matsubara frequency while the sum runs over all the bosonic Matsubara frequencies $\Omega_m=2m\pi/\beta$. 
The factor in square brackets in Eq.~(\ref{eq:sigmaell}), which depends on the angle $\theta_{{\bm k},{\bm k}+{\bm q}}$ between ${\bm k}$ and ${\bm k}+{\bm q}$, captures the dependence of Coulomb scattering on the relative chirality $s$ of the interacting electrons. 
The Green's function $ G^{(0)}_{\ell, s}({\bm k},i\omega) = 1/[i\omega - \xi_{\ell, s}({\bm k})]$ describes the free propogation of 
states with wavevector ${\bm k}$, Dirac energy $ \xi_{\ell, s}({\bm k})=s v k-\mu_\ell$ (relative to the chemical potential $\mu_\ell$ of the $\ell$-th layer) and chirality $s=\pm$. Note that in the r.h.s. of Eq.~(\ref{eq:sigmaell}) there are no terms involving products of the form 
$W_{\ell\ell'}G^{(0)}_{\ell',s}$ with $\ell' \neq \ell$. The reason is that, due to the absence of inter-layer tunneling, bare 
propagators are diagonal in the layer index: in the diagrammatic language (see Fig.~8.17 in Ref.~\onlinecite{Giuliani_and_Vignale}), screened interaction $W_{\ell\ell'}$ wavy lines closed on a bare propagator cannot begin on layer label $\ell$ and terminate on layer label $\ell' \neq \ell$.

After analytic continuation from imaginary to real frequencies, $i \omega \to 
\omega + i \eta$, the renormalized Fermi velocity $v^\star_\ell$ for the quasiparticles in the $\ell$-th layer can be expressed in terms of the wavevector and 
frequency derivatives of the retarted self-energy $\Sigma^{\rm ret}_\ell({\bm k},\omega)$ evaluated at the Fermi surface ($k=k_{{\rm F}, \ell}$) and at the quasiparticle pole $\omega=\xi_{\ell, +}({\bm k})$:
\begin{equation}\label{eq:v_star_dyson}
\frac{v^\star_\ell}{v}
=\frac{\displaystyle 1+(v)^{-1}\left.\partial_k \Re e \Sigma^{\rm ret}_\ell({\bf k},\omega)\right|_{k=k_{{\rm F}, \ell};\omega=0}}{1-\left.
\partial_{\omega} \Re e \Sigma^{\rm ret}_\ell({\bm k},\omega)\right|_{k=k_{{\rm F}, \ell};\omega=0}}~.
\end{equation}
As explained elsewhere~\cite{Giuliani_and_Vignale,polini_ssc_2007,asgari_prb_2005}, the derivatives in Eq.~(\ref{eq:v_star_dyson}) can both be expressed in terms of integrals along the imaginary frequency axis.

The ground-state energy of MLG can be easily evaluated from ${\hat {\cal H}}$. 
Apart from the trivial kinetic energy contribution $\delta \varepsilon_{\rm kin}$ related to the first term in ${\hat {\cal H}}$, it contains intra- and inter-layer interaction energy contributions. To evaluate the interaction energy we have followed a familiar strategy~\cite{Giuliani_and_Vignale,barlas_prl_2007} by combining the coupling-constant integration algorithm with the fluctuation-dissipation theorem~\cite{Giuliani_and_Vignale,barlas_prl_2007}. Following Ref.~\onlinecite{barlas_prl_2007}, we choose the total energy of undoped MLG (Fermi energy at the neutrality point in all layers) as our zero of energy. We separate the contribution that is first order in $\alpha_{\rm ee}$, the exchange energy, from the higher order contributions conventionally referred to in electron-gas theory as the correlation energy. 
Generalizing the theory for SLG~\cite{barlas_prl_2007} to a multicomponent system, we end up with rather cumbersome expressions for the exchange $\delta \varepsilon_{\rm x}$ and RPA correlation energies $\delta \varepsilon^{\rm RPA}_{\rm c}$ of MLG (per electron) measured from the reference undoped system. Explicit expressions for the $N=2$ case have been reported in Appendix~\ref{appendix}. 
Because these expressions involve only imaginary axis frequency integrals they are easy to evaluate accurately by quadrature. 

The determination of the electro-chemical equilibrium, discussed in more detail below, requires knowledge of exchange and correlations contributions to the chemical potentials $\mu_\ell$.  These can be easily calculated from
\begin{equation}
\mu_\ell  = \frac{\partial (n \delta \varepsilon_{\rm tot})}{\partial n_\ell}~,
\end{equation} 
where $n = n_1 + n_2 + \dots n_N$ is the total density and $\delta \varepsilon_{\rm tot} = \delta \varepsilon_{\rm kin} + \delta \varepsilon_{\rm x} + \delta \varepsilon^{\rm RPA}_{\rm c}$ is the total energy per electron, a function of $n_1, \dots, n_N$.

\section{Electro-chemical equilibrium}
\label{sect:electrochemicalequilibrium}

The equilibrium density distribution across a multilayer 
is achieved when the electro-chemical potentials in each distinct electronic region are identical. 
For a $N$-layer graphene system grown on a (SiC, say) substrate we identify $N+1$ electronic regions; the additional system is a buffer layer (not shown in Fig.~\ref{fig:one}) between the bulk substrate 
and the graphene layers which is positively charged as a consequence of 
charge transfer to the graphene system. The equilibrium densities in all the $N+1$ layers are 
determined by $N+1$ equations which express the discontinuity of the electric dispacement across charged layers.
With the notation introduced in Fig.~\ref{fig:one}, Gauss's law implies that for $\ell=N, N-1, \dots, 1$ (from bottom to top)
\begin{equation}\label{eq:gauss}
\epsilon_{\ell + 1} E_{\ell + 1} - \epsilon_{\ell} E_{\ell} = - 4 \pi e \sigma_{\ell}~,
\end{equation}
where $\sigma_{\ell}$ is the areal electron density in the $\ell$-th layer.  We can view $E_{1}$ as a quantity which is 
determined by a gate voltage.  As long as the distance $d_{\rm TB}$ from the top gate electrode to the MLG system is 
large, ``chemical" potential contributions are negligible so that $e E_{1} d_{\rm TG} = V_{\rm TG}$. 
We can write down an equation similar to (\ref{eq:gauss}) for the buffer layer, 
$\epsilon_{N+2} E_{N+2} - \epsilon_{N+1} E_{N+1} = + 4 \pi e \sigma_{\rm b}$. Again we can view $E_{N+2}$ as an experimentally controllable parameter (fixed by a bottom gate, $eE_{N+2} d_{\rm BG}= V_{\rm BG}$).  Here $\sigma_{\rm b}$ is the density of {\it positive} charges in the buffer layer.  This model assumes that the substrate is essentially
insulating.  In the absence of gates or unintended dopants that might create electric fields, we would normally
expect $E_{1}=E_{N+2}=0$.  In this case we obtain the charge neutrality condition: 
$\sum_{\ell = 1}^{N} \sigma_{\ell} = \sigma_{\rm b}$. 
In general $\sigma_{\rm b}$ should not be considered a free parameter.  It is determined by the total 
graphene charge and the difference between the two experimentally fixed electric fields $E_{1}$ and $E_{N+2}$. 
For MLG samples grown on SiC $\sigma_{\rm b}$ depends on growth parameters in a way which is at 
present not well understood.  

As explained above, our convention for the zero-of-energy of {\it chemical} energy of each layer is that is is zero at the the Dirac point,
{\it i.e.}, at electrical neutrality.  It is convenient to also choose an explicit global zero for the electric potential,
which we take to be its value on the top layer. Given the charge densities, the electrical potentials in each layer can be conveniently calculated 
iteratively starting from the top layer.  For $\ell =1, 2, \dots, N-1$ we have that
\begin{equation}
V_{\ell+1}=V_\ell+ e E_{\ell+1} d_{\ell+1,\ell}~. 
\end{equation}
Here $d_{\ell+1,\ell}$ is the separation between the $\ell+1$-th and the $\ell$-th layer. 
The condition for equilibrium between a graphene layer and the layer 
below it is: 
\begin{equation}\label{eq:equilibrium}
\mu_{\ell+1}+V_{\ell+1}= \mu_{\ell}+V_{\ell}~.
\end{equation}
We need one more equation to fix the densities and that is the equilibrium condition between the 
bottom layer and the buffer: $V_{\rm b} + \phi = V_{N} + \mu_{N}$. Here  
$\phi$ represents the microscopic physics which causes electrons to spill out of the buffer layer (where they are likely
poorly bonded).  It is known that $\phi$ is sensitive to the arrangement of carbon atoms in the buffer layer.
If all of the graphene layers are negatively charged there will be a large 
electric field between the $N$-th layer and the buffer layer whose sign will tend to repel electrons,
{\it i.e.} to make $V_{\rm b} < V_{N}$.   The value of $\phi$ must therefore be positive and it should be
chosen to match experimental results for the total charge of all layers in a MLG systems.
We assume that $\phi$ is fixed once a sample has been prepared, {\it i.e.} that it is not influenced by 
gate voltages, external magnetic fields, or other parameters that are routinely used to 
alter the electrical properties of two dimensional electron systems.  

\section{Numerical results and discussion}
\label{sect:numericalresults}

We now turn to the presentation of detailed illustrative 
numerical results for double-layer graphene (DLG). 
The bare intra- and inter-layer Coulomb interactions are influenced by the layered dielectric environment.
For the $N=2$ case, a routine electrostatics calculation~\cite{bare_interactions} implies that the Coulomb interaction in the $\ell =1$ (top) layer is given by
\begin{equation}\label{eq:v11}
V_{11}(q) = \frac{4\pi e^2}{q D(q)} [ (\epsilon_2 + \epsilon_3) e^{qd} + 
 (\epsilon_2 - \epsilon_3) e^{-qd}]~,
\end{equation}
where
\begin{equation}\label{eq:denominator}
D(q) =  [(\epsilon_1 + \epsilon_2) (\epsilon_2 + \epsilon_3) e^{qd} + 
(\epsilon_1 - \epsilon_2) (\epsilon_2 - \epsilon_3) e^{-qd} ] 
\end{equation}
and $d$ is a shorthand notation for the inter-layer distance $d_{2,1}$. 
The Coulomb interaction in the bottom layer, $V_{22}(q)$, can be simply obtained from $V_{11}(q)$ 
by interchanging $\epsilon_3 \leftrightarrow \epsilon_1$. Finally, the inter-layer Coulomb interaction is given by
\begin{equation}\label{eq:v12} 
V_{12}(q) = V_{21}(q) = \frac{8\pi e^2}{q D(q)}~\epsilon_2~.
\end{equation}
In the case $N=2$ the matrix equation (\ref{eq:matrix-form}) can be easily inverted and the screened potentials 
${\bm W}$ in Eq.~(\ref{eq:e-e}) can be written in a particularly compact form~\cite{zheng_prb_1994}:
\begin{equation}\label{eq:W11}
W_{11}(q,\omega) = \frac{V_{11}(q)+[V^2_{12}(q)-V_{11}(q)V_{22}(q)]\chi^{(0)}_2(q,\omega)}{\varepsilon(q,\omega)}
\end{equation}
and
\begin{equation}\label{eq:W12}
W_{12}(q) = \frac{V_{12}(q)}{\varepsilon(q,\omega)}~.
\end{equation}
The expression for $W_{22}(q)$ can be obtained from Eq.~(\ref{eq:W11}) by interchanging $1 \leftrightarrow 2$. In Eqs.~(\ref{eq:W11})-(\ref{eq:W12}) we have introduced the dielectric function
\begin{eqnarray}
\varepsilon(q,\omega) &=& [1-V_{11}(q) \chi^{(0)}_1(q,\omega)][1-V_{22}(q)\chi^{(0)}_2(q,\omega)] \nonumber\\
&-& V^2_{12}(q)\chi^{(0)}_1(q,\omega)\chi^{(0)}_2(q,\omega)~.
\end{eqnarray}

Exchange and correlation energies in MLG systems depend both on interactions on the Fermi wavelength scale
which influence correlations between carriers and
on interactions at shorter length scales which influence correlation with the Dirac sea background.  
This contrasts sharply with the case of ordinary two-dimensional 
electron gases for which many-body properties depend only on interactions on the Fermi wavelength scale.
Because MLG layers are separated by atomic length scales and carrier densities are always small on atomic scales
$k_{{\rm F},\ell} d$ is typically small.  Taking the limit $q d \to 0$ in Eqs.~(\ref{eq:v11})-(\ref{eq:v12}) we find that the 
typical carrier-carrier interactions are layer independent with effective dielectric constant
$(\epsilon_1+\epsilon_3)/2$.  Similarly, taking the limit $q d \to \infty$ we find that carrier-background 
interactions are approximately independent in different layers, and that they have an effective dielectric constant 
determined mostly by immediately adjacent layers.   

With these explicit expressions in our hands we are in the position to calculate the quasiparticle self-energies $\Sigma_\ell$ and velocities $v^\star_\ell$, the exchange $\delta \varepsilon_{\rm x}$ and RPA correlation $\delta \varepsilon^{\rm RPA}_{\rm c}$ energies, and to solve the electro-chemical equilibrium problem outlined above. For the presentation of the numerical results for the double-layer system
we introduce the density polarization $\zeta = (n_2 - n_1)/n$: $\zeta = 1$ when the carrier density is non-zero only in the bottom layer, 
while $\zeta = 0$ when the two layers have identical carrier densities. 

\begin{figure}[t]
\centering
\begin{tabular}{c}
\includegraphics[width=0.90\linewidth]{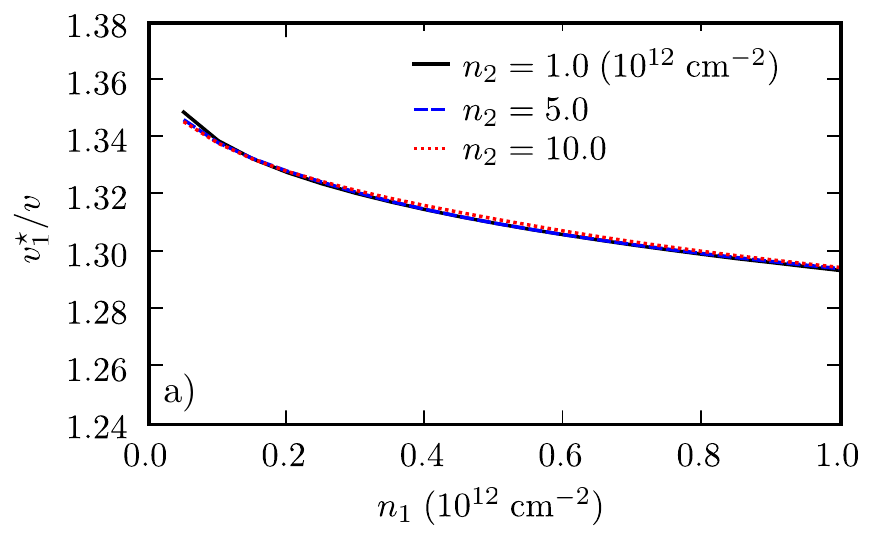}\\
\includegraphics[width=0.90\linewidth]{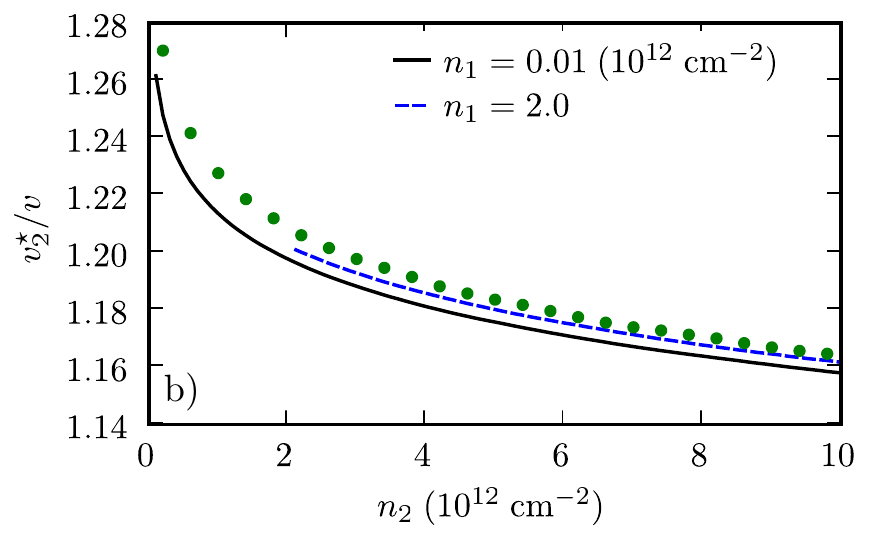}
\end{tabular}
\caption{(Color online) Quasiparticle velocities in double-layer graphene on SiC [$\epsilon_1 = \epsilon_2 = 1.0$ and $\epsilon_3 = 6.6$ (SiC dielectric constant)]. The data shown in this figure have been obtained by setting $d = 3.35$~\AA~and $\alpha_{\rm ee} = 2.2$. 
Panel a) Quasiparticle velocity $v^\star_1$ in the top layer (in units of the bare velocity $v$) as a function of the density $n_1$ in that layer (in units of $10^{12}~{\rm cm}^{-2}$), for different values of the density $n_2$ in the opposite layer. 
Panel b) Quasiparticle velocity $v^\star_2$ in the bottom layer (in units of the bare velocity $v$) as a function of the density $n_2$ in that layer (in units of $10^{12}~{\rm cm}^{-2}$), for different values of the density $n_1$ in the opposite layer.
Circles label data for the quasiparticle velocity in single-layer graphene~\cite{polini_ssc_2007} on SiC. 
\label{fig:two}}
\end{figure}

In Fig.~\ref{fig:two} we present results for the quasiparticle velocity enhancements in the low and high density layers, $v^\star_1/v$ and $v^\star_2/v$, as functions of their carrier density for several different values of the density in the opposite layer.  The quasiparticle velocity in a single-layer has a large enhancement due mainly~\cite{polini_ssc_2007,diracgaspapers} to exchange interactions between the carriers and the Dirac sea.  Carrier-carrier interactions yield positive contributions to both the wavevector and frequency dependence of the self-energy, which partially cancel as contributions to the renormalized quasiparticle velocity much as they do in the ordinary two-dimensional electron gas.  The end result is that the enhanced velocity due to exchange interactions with the Dirac sea largely survives, especially
at low carrier densities.  The parameters of this calculation were chosen with DLG on SiC in mind.  The general result is that enhanced quasiparticle velocity again survives, with quantitative changes due to electron-electron interactions.  Two aspects of these results may appear counter-intuitive at first glance. First of all, we see that inter-layer interactions are important even when the carrier density in remote layers is zero.  Because the Dirac bands are gapless and because both bands have $\pi$-orbital character, their wavevector and frequency dependent polarization function is important in forming the dynamically screened Coulomb interaction even in the absence of carriers.  Secondly, we notice that increasing the density of a remote layer does not necessarily weaken the screened interaction in the layer of interest.  The origin of this property can be traced to the second term in Eq.~(\ref{eq:W11}) which is not intuitive and captures the mutual screening response of double-layers including inter-layer interaction effects. 
%

%%%%%%%%%%%%%%
%%%%%%%%%%%%%%

\begin{figure}[t]
\centering
\includegraphics[width=0.90\linewidth]{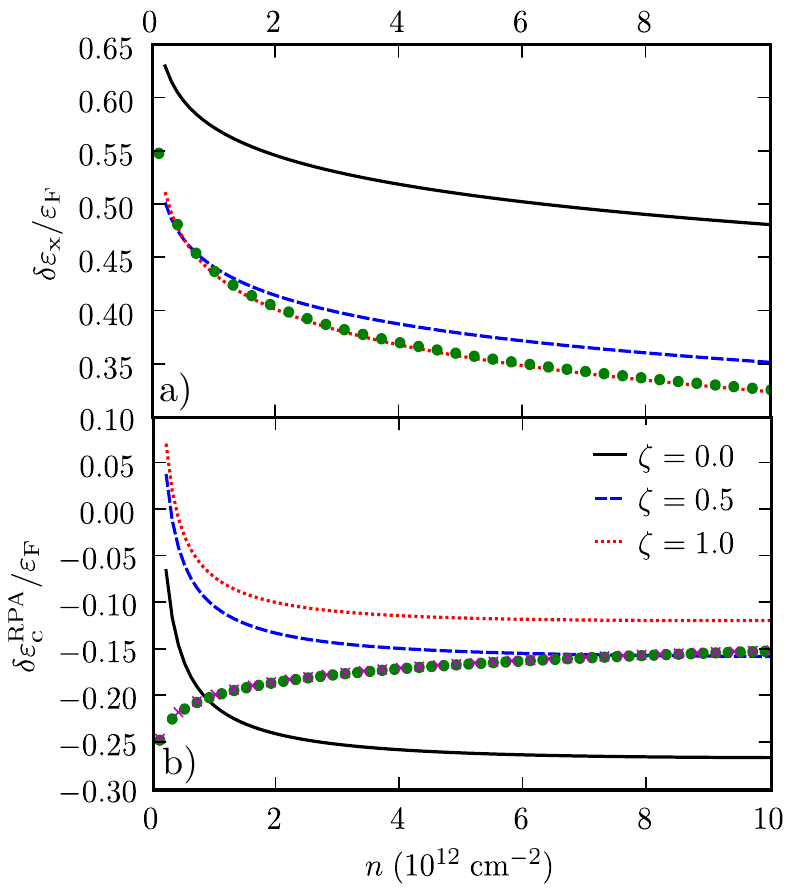}
\caption{(Color online) Interaction energies in double-layer graphene on SiC, measured from the reference undoped system. 
All parameters are the same as in Fig.~\ref{fig:two}. Panel a) 
The exchange energy $\delta \varepsilon_{\rm x}$ in units of the Fermi energy $\varepsilon_{\rm F}$ 
as a function of $n$ (in units of $10^{12}~{\rm cm}^{-2}$) and 
for different values of $\zeta$. Panel b) Same as in the top panel but for the RPA correlation energy $\delta \varepsilon^{\rm RPA}_{\rm c}$. Crosses label the RPA correlation energy for $\zeta=1$ if the inter-layer interaction $V_{12}$ is set to zero. 
Circles label the exchange and RPA correlation energies in single-layer graphene~\cite{barlas_prl_2007} on SiC. Note that the correlation energy obtained neglecting inter-layer interactions (crosses) is practically equal to the single-layer graphene result (circles).\label{fig:three}}
\end{figure}

In Fig.~\ref{fig:three} we present numerical results for the DLG exchange and RPA correlation energies per carrier, $\delta \varepsilon_{\rm x}$ and $\delta \varepsilon^{\rm RPA}_{\rm c}$.
In this case we plot energies as a function of total carrier density for several different layer polarizations in order to illustrate 
a relationship to well known dependences on spin-polarization that we discuss below.
Exchange energies are positive because~\cite{barlas_prl_2007,polini_prb_2008} they are calculated relative to zero carrier density using 
the Dirac point self-energy of this limit as the zero of energy.  The increase in exchange energy with 
carrier density in graphene has the physical consequence that corrections to the RPA 
are expected~\cite{polini_prb_2008} to enhance screening, instead of weakening it as in an ordinary two-dimensional electron gas.
Notice that the exchange energy of DLG is larger than that of SLG.  The two are equal for $\zeta=1$ because the 
exchange energy depends only on intra-layer interactions [see Eq.~(\ref{eq:exchangeenergy}) in Appendix~\ref{appendix}]. 
The larger exchange energy in the DLG case, in which 
carriers are separated between two different layers, has an origin similar to the well known increase in the 
exchange energy of an electron gas when two different spin states are occupied.  The correlation energy, which is 
negative, is dramatically larger in DLG and strongly influenced by inter-layer interactions.  These correlation
contributions to the energy suggest that exchange-only approximations overestimate the degree to which 
screening is enhanced by beyond-RPA corrections.
When correlation are included, the interaction energy preference for unequal partitioning of density between layers 
is also strongly reduced.  This result has an origin similar to the well known result that exchange-only theories 
badly overestimate the tendency of interactions to favor enhanced spin polarization in ordinary electron gases.

%%%%%%%%%%%%%%
%%%%%%%%%%%%%%

\begin{figure}[t]
\centering
\includegraphics[width=0.90\linewidth]{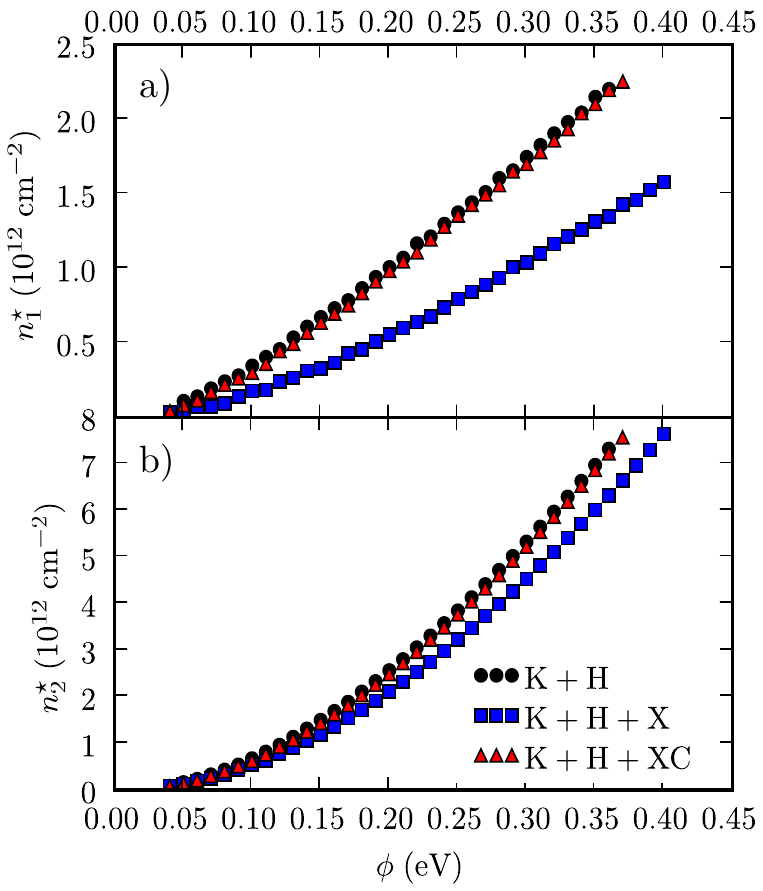}
\caption{(Color online) Electro-chemical equilibrium for double-layer graphene on SiC in the absence of top and bottom gates ($E_1=E_4=0$). 
All parameters are the same as in Figs.~\ref{fig:two}-\ref{fig:three}. The distance between buffer and bottom layers is equal to $3.35$~\AA.
Panel a) Equilibrium density in the top layer, $n^\star_1$ (in units of $10^{12}~{\rm cm}^{-2}$), as a function of the chemical potential $\phi$ (in ${\rm eV}$) in the buffer layer. Here we have reported results obtained (i) neglecting intra- and inter-layer e-e interactions (circles), (ii) including e-e interactions at the exchange-only level (squares), and (iii) including both exchange and correlations (triangles). Panel b) Same as in panel a) but for the bottom layer. Note how $n^\star_2 > n^\star_1$.\label{fig:four}}
\end{figure}

Finally, in Fig.~\ref{fig:four} we present a typical result for the dependence of the 
equilibrium densities in top and bottom layers, $n^\star_1$ and $n^\star_2$, on the work function of the buffer layer. 
We clearly see that the equilibrium density in the top layer $n^\star_1$ is substantially smaller than the density in the bottom layer, the ratio $n^\star_2/n^\star_1$ changing between $1.5$ and $3.5$ in the range of of $\phi$ values explored in this figure. Quite importantly, note that including e-e interactions at the exchange-only level severely underestimates the values of the equilibrium densities, especially so in the top layer.
Including correlation effects reduces the energetic cost of adding electrons to the graphene layers.  In fact 
we find that the increase in exchange energy relative to the Dirac point exchange energy and the decrease in interaction
energy due to correlations among the carriers partially compensate, leading to densities in the graphene layers
close to those implied by a Hartree theory which completely ignores exchange and correlation effects.  
This finding is surprising in comparison to the ordinary two-dimensional electron gas case, in which 
exchange and correlation effects always lower the energy cost of adding carriers and 
increase charge transferred from an electron reservoir.
%%%%%%%%%%%%%%
%%%%%%%%%%%%%%

\section{Summary and conclusions}
\label{sect:summary}

In summary, we have presented a theoretical scheme, based on the random phase approximation and on the ``${\rm G}_0{\rm W}$" theory, to deal with electron-electron interactions in multi-layer graphene 
Fermi liquid systems consisting in which hybridization is suppressed by relative rotations of honeycomb lattices. We have presented numerical results for the specific case of a double layer, explicitly demonstrating that inter-layer Coulomb interactions substantially alter the properties of the quasiparticles in this system, with respect to those of isolated single-layer graphene. We have also solved numerically the electro-chemical equilibrium problem, finding that it is crucial to incorporate correlation effects when going beyond a free-electron description of this system.
When both effects are included, interactions play a major role in lowering the interaction energy of MLG systems.  Because of strong 
inter-layer carrier-carrier interactions, the total interaction energy is relatively insensitive to charge distribution among the layers,
which is well approximated by a theory that includes only kinetic and electrostatic energies.    
The unusual positive exchange energy of bilayer graphene acts to suppress charge transfer from the buffer layer.  
Surprisingly, when both exchange and correlation is included corrections to the Hartree theory of charge transfer from the MLG buffer layer are quite small, at least in the $N=2$ MLG system.  In analysing the physics of larger $N$ MLG systems, it will be essential to recognize that layers contribute to the systems's exchange and correlation energy even when their carrier density vanishes.  Because carrier-carrier interactions are weakly layer dependent, MLG systems might provide an interesting approximate realization of a $SU(N)$ electron gas model in which $1/N$ expansions are quantitatively useful.  This might provide an interesting attack on a limit in which corrections to the ${\rm G}_0{\rm W}$ theory 
can be explored by a rigorous semi-classical theory.     
   
\acknowledgments

A.H.M. acknowledges support from the Welch Foundation and from the SWAN NRI program.
We thank C. Berger, M. Fogler, and M. Potemski for useful conversations.

\appendix

\section{Explicit expressions for the exchange and RPA correlation energies of double-layer graphene}
\label{appendix}

For the sake of completeness, in this Appendix we report the expressions we have used to compute the exchange and RPA correlation energies of DLG. In what follows all symbols with a bar over them denote dimensionless quantities. We have scaled all wavevectors with the Fermi wavevector $k_{\rm F} = \sqrt{\pi n}$ evaluated at the total density $n$, all frequencies with the Fermi energy $\varepsilon_{\rm F} = v k_{\rm F}$, the Lindhard response functions $\chi^{(0)}_\ell$ with the massless Dirac fermion density-of-states $g \varepsilon_{\rm F}/(2\pi v^2)$, and the bare Coulomb potentials $V_{\ell\ell'}$ with $2\pi e^2/ k_{\rm F}$. Here $g =4$ is the usual spin-valley degeneracy factor.

It is easy to prove that the exchange energy can be written as
\begin{eqnarray}\label{eq:exchangeenergy}
\delta \varepsilon_{\rm x}
&=-&\frac{\varepsilon_{\rm F}\alpha_{\rm ee}}{\pi} \left\{\int_0^{+\infty}d{\bar q}~{\bar q}{\bar V}_{11}({\bar q})
\int_0^{+\infty}d{\bar \Omega}~\delta{\bar \chi}^{(0)}_{1}({\bar q},i{\bar \Omega})\right. \nonumber\\
&+&\int_0^{+\infty}d{\bar q}~ {\bar q}{\bar V}_{22}({\bar q})
\int_0^{+\infty}d{\bar \Omega}~\left.\delta {\bar \chi}^{(0)}_{2}({\bar q},i{\bar \Omega})\right\}~,
\end{eqnarray}
while the RPA correlation energy as
\begin{widetext}
\begin{eqnarray}\label{eq:correlationenergy}
\delta \varepsilon_{\rm c}&=&\frac{\varepsilon_{\rm F}\alpha_{\rm ee}}{\pi} 
\left\{\int_0^{+\infty} d{\bar q}~{\bar q}{\bar V}_{11}({\bar q})
\int_0^{+\infty}d{\bar \Omega}~\left[\delta\Phi_{11}({\bar q}, i{\bar \Omega})
+\delta{\bar \chi}^{(0)}_{1}({\bar q}, i{\bar \Omega})\right]\right.\nonumber\\
&+&\left.\int_0^{+\infty} d {\bar q}~{\bar q}{\bar V}_{22}({\bar q})
\int_0^{+\infty}d{\bar \Omega}\left[\delta\Phi_{22}({\bar q}, i{\bar \Omega})
+\delta{\bar \chi}^{(0)}_{2}({\bar q}, i{\bar \Omega})\right] + 2\int_0^{+\infty} d{\bar q}~{\bar q}{\bar V}_{12}({\bar q})
\int_0^{+\infty}d{\bar \Omega}~\delta\Phi_{12}({\bar q},i{\bar \Omega})\right\}~.
\end{eqnarray}
\end{widetext}
In Eq.~(\ref{eq:correlationenergy}) we have introduced the quantities (for reasons of space we will omit to write explicitly the dependence of the following expressions on the variables $q$ and $i\Omega$)
\begin{widetext}
\begin{eqnarray}\label{eq:phi11}
\Phi_{11} &=& -\frac{1}{(g\alpha_{\rm ee})^2 {\bar \chi}_{2}^{(0)}
({\bar V}_{11}{\bar V}_{22} - {\bar V}^2_{12})} 
\left\{\frac{2L - (g\alpha_{\rm ee}){\bar V}_{22} \bar{\chi}_{2}^{(0)} (K+\sqrt\Delta)}{2\sqrt\Delta}
\ln\left|\frac{2L - K - \sqrt\Delta}{-K-\sqrt \Delta}\right|\right.\nonumber\\
&-&\left. \frac{2L - (g\alpha_{\rm ee}){\bar V}_{22}{\bar \chi}^{(0)}_{2}(K-\sqrt\Delta)}{2\sqrt\Delta}\ln\left|\frac{2L-K+\sqrt\Delta}{-K+\sqrt \Delta}\right|\right\}
\end{eqnarray}
and
\begin{equation}\label{eq:phi12}
\Phi_{12}  = -\frac{{\bar V}_{12}}{(g\alpha_{\rm ee})({\bar V}_{11}{\bar V}_{22}
-{\bar V}^2_{12})}\left\{\frac{K+\sqrt{\Delta}}{2\sqrt{\Delta}}\ln\left|\frac{2L-K-\sqrt{\Delta}}{-K-\sqrt{\Delta}}\right|
-\frac{K-\sqrt{\Delta}}{2\sqrt{\Delta}}\ln\left|\frac{2L-K+\sqrt{\Delta}}{-K+\sqrt{\Delta}}\right|
\right\}~,
\end{equation}
\end{widetext}
where
\begin{equation}\label{eq:chiud}
\Delta = (g\alpha_{\rm ee})^2 \left[({\bar V}_{11}{\bar \chi}^{(0)}_1 - {\bar V}_{22}{\bar \chi}^{(0)}_2)^2 
+ 4 {\bar V}^2_{12} {\bar \chi}^{(0)}_1 {\bar \chi}^{(0)}_2\right]~,
\end{equation}
\begin{equation}
L = (g\alpha_{\rm ee})^2\left({\bar V}_{11}{\bar V}_{22}{\bar \chi}^{(0)}_1{\bar \chi}^{(0)}_2 
- {\bar V}^2_{12}{\bar \chi}^{(0)}_1 {\bar \chi}^{(0)}_2\right)~,
\end{equation}
and
\begin{equation}
K = g\alpha_{\rm ee}\left({\bar V}_{11}{\bar \chi}^{(0)}_1 + {\bar V}_{22}{\bar \chi}^{(0)}_2\right)~.
\end{equation}
The expression for $\Phi_{22}$ can be obtained from Eq.~(\ref{eq:phi11}) by interchanging $1 \leftrightarrow 2$.

Finally, we remark that all the quantities which in Eqs.~(\ref{eq:exchangeenergy})-(\ref{eq:correlationenergy}) 
are preceded by the sign ``$\delta$" indicate quantities defined by differences between doped and undoped values: for example
\begin{equation}
\delta {\bar \chi}^{(0)}_\ell ({\bar q}, i {\bar \Omega}) 
=  {\bar \chi}^{(0)}_\ell({\bar q}, i {\bar \Omega}) 
- \left. {\bar \chi}^{(0)}_\ell({\bar q}, i {\bar \Omega})\right|_{k_{{\rm F}, \ell} =0}~,
\end{equation}
and
\begin{equation}
\delta \Phi_{ij}({\bar q}, i{\bar \Omega}) = \Phi_{ij}({\bar q}, i{\bar \Omega}) - 
\left.\Phi_{ij}({\bar q}, i{\bar \Omega})\right|_{k_{{\rm F}, \ell} =0}~.
\end{equation}
Note that the dimensionless Lindhard function in the undoped limit is the same for every layer and is given by
\begin{equation}
\left. {\bar \chi}^{(0)}_\ell({\bar q}, i {\bar \Omega})\right|_{k_{{\rm F}, \ell} =0} = -\frac{\pi}{8}\frac{{\bar q}^2}{\sqrt{{\bar q}^2 + {\bar \Omega}^2}}~.
\end{equation}


\begin{thebibliography}{77}
\bibitem{bare_interactions}
	The Coulomb potentials $V_{\ell \ell'}(q)$ can be 
	determined by solving the Poisson equation $\nabla^2 \varphi({\bm \rho}, z) = 0$ for the electrical 
	potential $\varphi({\bm \rho}, z)$ in cylindrical coordinates.
	The solution of the Poisson equation in all the $N+1$ regions can be written as $\varphi_\ell({\bm \rho}, z) = 
	\exp{(i {\bm q} \cdot {\bm \rho})}[\alpha_\ell \exp(q z) +\beta_\ell \exp(-qz)]$ and depends on a total of $2(N+1)$ constants.
	Two are fixed by imposing boundary conditions $\varphi({\bm \rho}, z \to \pm \infty)=0$ at infinity. The remaining $2N$
	constants are fixed by the continuity of the potential $\varphi$ across each layer 
	and by imposing that the electric displacement 
	has a discontinuity. Let $D_{\ell} = \epsilon_{\ell} E_{\ell}$ be the electric displacement 
	above the $\ell$-th layer and $\epsilon_{\ell}$ be the dielectric constant in this region of space ($E= - \partial_z \varphi$). 
	Then one has to impose that $D_{\ell+1} - D_{\ell} = 4\pi \sigma_{\ell}$ for $\ell = N \dots 1$, 
	where $\sigma_\ell = - e$ is the Fourier transform of $-e \delta^2({\bm \rho})\delta(z - (\ell-1) d)$.
\end{thebibliography}
\end{document}